\documentclass{aa}  
\usepackage{tikz}
\usepackage{physics}
\usepackage{amsmath}
\usepackage{float} 
\usepackage{txfonts}
\usepackage[colorlinks=true,citecolor=blue,urlcolor=blue]{hyperref}
\usepackage{threeparttable}
\usepackage{graphicx}
\usepackage{txfonts}
\usepackage[dvipsnames]{xcolor}

\begin{document}

   \title{Evidence for multiple crossings and stripping of Gaia-Enceladus/Sausage across the Milky Way}

   \author{L. Berni\inst{1, 2}\fnmsep\thanks{\email{leda.berni@unifi.it}},
           M. Palla\inst{1,3, 4},
           L. Magrini\inst{1}
           \and
           L.Spina\inst{1,5}
          }

   \institute{INAF - Osservatorio Astrofisico di Arcetri, Largo E. Fermi 5, 50125, Firenze, Italy\
         \and
             Dipartimento di Fisica e Astronomia, Università degli Studi di Firenze, Via Sansone 1, 50019, Sesto Fiorentino, Italy\
        \and Dipartimento di Fisica e Astronomia "Augusto Righi", Alma Mater Studiorum, Università di Bologna, Via Gobetti 93/2, 40129, Bologna, Italy
        \and INAF, Osservatorio di Astrofisica e Scienza dello Spazio, Via Gobetti 93/3, 40129, Bologna, Italy;
        \and INAF - Padova Observatory, Vicolo dell’Osservatorio 5, 35122 Padova, Italy
    }

  \abstract
   {The accretion of Gaia-Enceladus/Sausage (GES) onto the Milky Way (MW) is one of the most prominent features of the Galactic halo revealed by the combination of the {\it Gaia} satellite and large spectroscopic surveys. This massive accretion largely contributes to the local stellar halo mass and was significant enough to alter the formation history and the morphology of the MW.
   }
   {In this work, we aim to analyse the selection of stars previously identified as belonging to GES with different kinematics and chemical properties to test the hypothesis of a two-phase accretion event.}
   {We apply several statistical tests to assess the significance of the separation between the two populations in GES. We then employ galactic chemical evolution models to investigate the origin of the chemical differences encountered in the analysis.}
   {We confirm the presence of two distinct populations, with consistently different dynamical and chemical properties. The low energy population seems to show higher overall abundances, whereas the high-energy one may be more metal-poor. We attribute this difference to the presence of at least two separate populations of stars within Gaia-Enceladus, likely associated with the innermost (low-energy) and outermost (high-energy) regions of the progenitor. The adopted models successfully reproduce the patterns in metallicity and [$\alpha$/M] distributions in an inside-out scenario.}
   {Our analysis supports the presence of a former metallicity gradient in Gaia-Enceladus, and reinforces the interpretation of its accretion as a multi-passage event through the Milky Way disc.}

\titlerunning {Multiple crossings of GES}
\authorrunning{Berni, Palla et al.}
   \maketitle

\section{Introduction}
The study and identification of stellar structures in the Milky Way (MW), made possible by large astrometric and spectroscopic surveys such as {\it Gaia} \citep{GaiaCollab2016}, the Apache Point Observatory Galactic Evolution Experiment  \citep[][hereafter APOGEE]{majewski2017apache}, and many others, have opened a new window on our understanding of the formation and evolution of our Galaxy.
It is now well established that a significant fraction, up to roughly 50\% of the stellar halo, did not form `in situ', i.e. within the MW galaxy, but was instead accreted from disrupted globular clusters (GCs) and dwarf galaxies \citep[see e.g.][]{Bell2008,helmi20}.

Theoretical models suggest that during accretion events, the orbital energy and angular momentum of stars are mostly conserved \citep{helmi2000mapping}. However, repeated dynamical interactions with the MW can strip stars from their original clusters or galaxies, making it challenging to identify individual stars from disrupted systems that are now mixed into the halo field.

Accreted structures can usually be distinguished from the `in situ' population through their distinct chemical abundance patterns or by their peculiar orbital dynamics \citep[see. e.g.][]{Horta2022}. 
Stars that form in the same molecular cloud or in the same region of a galaxy tend to share similar chemical abundances, while those born in different regions usually show distinct abundance patterns linked to their different star formation histories \citep[see, e.g.][]{Dropulic2025ApJ...990..162D}. 
Since photospheric abundances remain largely unchanged throughout a star’s lifetime for most elements, they provide a powerful means to link stars to the stellar population where they are born—a method known as chemical tagging.
Indeed, in the Galactic community the concept of weak chemical tagging is widely applied, aiming to distinguish different galactic populations with distinct chemical enrichment histories. In practice, chemical tagging is typically carried out using a combination of chemical, kinematic, and age information to robustly identify stellar populations with common origin \citep[e.g.][]{Freeman2002ARA&A..40..487F, Cheng2021MNRAS.506.5573C,  Bhattarai2024ApJ...977...70B,  Spina2025arXiv250918268S}.
Chemical tagging is particularly relevant, because over time, the tidal forces exerted by the MW can elongate globular clusters and dwarf galaxies into stellar streams, or even completely disrupt them, dispersing their stars into the diffuse halo field population. With the disruption, they mostly lose spatial and also dynamical coherence, but not the inherited chemical abundances.

Many studies have attempted to detect and characterize such substructures in the MW halo, often relying on clustering algorithms applied to the dynamical space \citep[see e.g.][]{Ding2025}, and, more recently, combining these with chemical abundance analyses (e.g. \citealt{chen2018chemodynamical, Lovdal2022, Ceccarelli2024, berni2025exploring}).
In this framework, the {\it Gaia} mission revolutionised our understanding of the MW’s accretion history by enabling the identification of one of its most significant merger events, the GES structure \citep{Helmi2018,Belokurov2018}.
GES is thought to be the remnant of a dwarf galaxy of about $10^9 M_\odot$ \citep[e.g.][]{Feuillet2020SkyMapper, Plevne2025ApJ...991..207P}, that merged with the MW about 10 Gyr ago \citep{Gallart2019Uncovering}.

This accretion event is believed to have been sufficiently massive to alter the formation history and morphology of the MW disc, possibly triggering the formation of the thick disc \citep{Funakoshi2025}. It has also been proposed as a potential cause for the emergence of a relatively metal-rich ([M/H] > -1) stellar population known as the Splash \citep[see e.g.][]{Belokurov2020}, though it remains possible that this mechanism did not contribute substantially to its formation \citep[see e.g.][]{amarante2020,Kisku2025}.
Finally, GES is estimated to have contributed approximately 20\% of the Milky Way’s present globular cluster system \citep[e.g.][]{Massari2019, Boldrini2025arXiv250613254B}, further highlighting its significance in the Galaxy’s early assembly history.
Given its massive nature and complex merger history, GES has become a natural laboratory for studying the processes of accretion, chemical evolution, and dynamical heating in the early Milky Way. Recent works have started to reveal that GES itself may not be a homogeneous structure, but rather composed of multiple subcomponents \citep[see, e.g.][]{skuladottir2025evidence, Buder2025arXiv251011284B, Carrillo2025arXiv250924705C}.

\citet{skuladottir2025evidence} presents compelling evidence that the accretion of GES onto the MW occurred in at least two passages through the Galactic disc, confirming the models from \citet{Naidu2021a} and \citet{han2022stellar}. Their analysis, however, relies mainly on the orbital energy of the stars to distinguish the two accreted components. Given the relatively high mass of GES, its stellar population is expected to be chemically diverse across its proto-disc, showing a possible evidence of radial abundance gradients.
As proposed in \citet{skuladottir2025evidence} and later confirmed with cosmological simulation by \citet{Buder2025arXiv251011284B} and by \citet{Carrillo2025arXiv250924705C}, the first disc passage stripped the outer, less metal enriched population of GES, characterized by higher orbital energies, while the second passage accreted a more metal-rich, lower-energy population corresponding to the inner regions of the progenitor galaxy.
This scenario is supported by slight chemical differences between the two groups both in data and simulations, particularly visible in the [$\alpha$/M]–[M/H] plane, where the lower-energy population shows systematically higher [$\alpha$/M] ratios at fixed metallicity.
In a recent paper, \citet{berni2025exploring} applies a machine-learning algorithm to a large stellar halo sample of the Milky Way observed within the APOGEE survey, successfully recovering the GES structure and two distinct subpopulations within it. Compared to previous works, this approach yields a more robust membership determination, as it accounts for both the chemical and dynamical properties of the stars.

In light of these recent results, we aim to reanalyse the GES sample from \citet{berni2025exploring} to test the hypothesis of a two-phase accretion event for GES through a detailed chemical characterisation and the application of statistical diagnostics.
The paper is organised as follows: the dataset, its chemical properties and the statistical methods used to distinguish the two stellar populations are described in Section \ref{sec:dataset}.
In Section \ref{sec:models}, we outline the results of the chemical enrichment models employed to test our hypothesis, and we compare them with the observations.
A summary of our findings and the main conclusions are given in Section \ref{sec:conclusions}.

\section{Dataset and chemical characterisation}
\label{sec:dataset}

To investigate the chemo-dynamical structure of the Gaia-Enceladus/Sausage system, we selected a sample of halo stars from the APOGEE survey. This section outlines the selection criteria and data preparation steps adopted to define the final sample used in our analysis and gives a chemical characterization of the data.

\subsection{Sample selection}

The Apache Point Observatory Galactic Evolution Experiment \citep[APOGEE;][]{majewski2017apache} is a spectroscopic survey conducted in the near infrared band using two 300-fiber cryogenic spectrographs operating at Apache Point Observatory (APO) in New Mexico, United States, and the 2.5 metre Irénée du Pont Telescope of Las Campanas Observatory (LCO) in Atacama de Chile.
The data release {\sc dr~17} provided parameters and chemical abundances for approximately 657,000 targets \citep{Abdurro2022ApJS..259...35A}.
The sample of stars used in this work is the result of a selection of the APOGEE dataset to identify stars belonging to the Galactic halo on the basis of their velocity and metallicity. In addition, we adopted several quality criteria to define our sample: signal-to-noise ratio (S/N) > 70, effective temperature (T$_{\rm eff}$) in the range 3500-6500 K, STARFLAG=0 and ASPCAPFLAG < 256, allowing the exclusion of stars with issues relating to parameter determination or spectral fitting problems.

A further selection was applied to identify stars belonging to the accreted structure of GES, based on the use of our machine learning algorithm called {\sc CREEK}, described in \citet{berni2025exploring}. This algorithm detects substructures in the Milky Way halo, leveraging its low density and old age, which allow it to retain signatures of past merger events. {\sc CREEK} has proven effective in re-identifying both globular clusters (GCs) and stellar streams.
The algorithm was applied to halo  stars with metallicity [M/H] < -1, to ensure a more stringent selection of halo members. 
Although this choice may cause us to miss the high-metallicity tail of the halo, it greatly reduces contamination from the disc population. Extending the metallicity range would increase disc contamination, thereby reducing the predictive power of the algorithm. {\sc CREEK} creates links between stars based on their dynamical similarity, learned through a siamese neural network \citep[see e.g.][]{Chicco2021}, and then groups them using a graph neural network autoencoder \citep[see. e.g.][]{Gori2005} based on their chemical abundances, specifically [M/H] and [$\alpha$/M]. The outcome is presented as a reachability plot (see \citealt{berni2025exploring} for details).

\begin{figure*}
    \centering
    \includegraphics[width=1\linewidth]{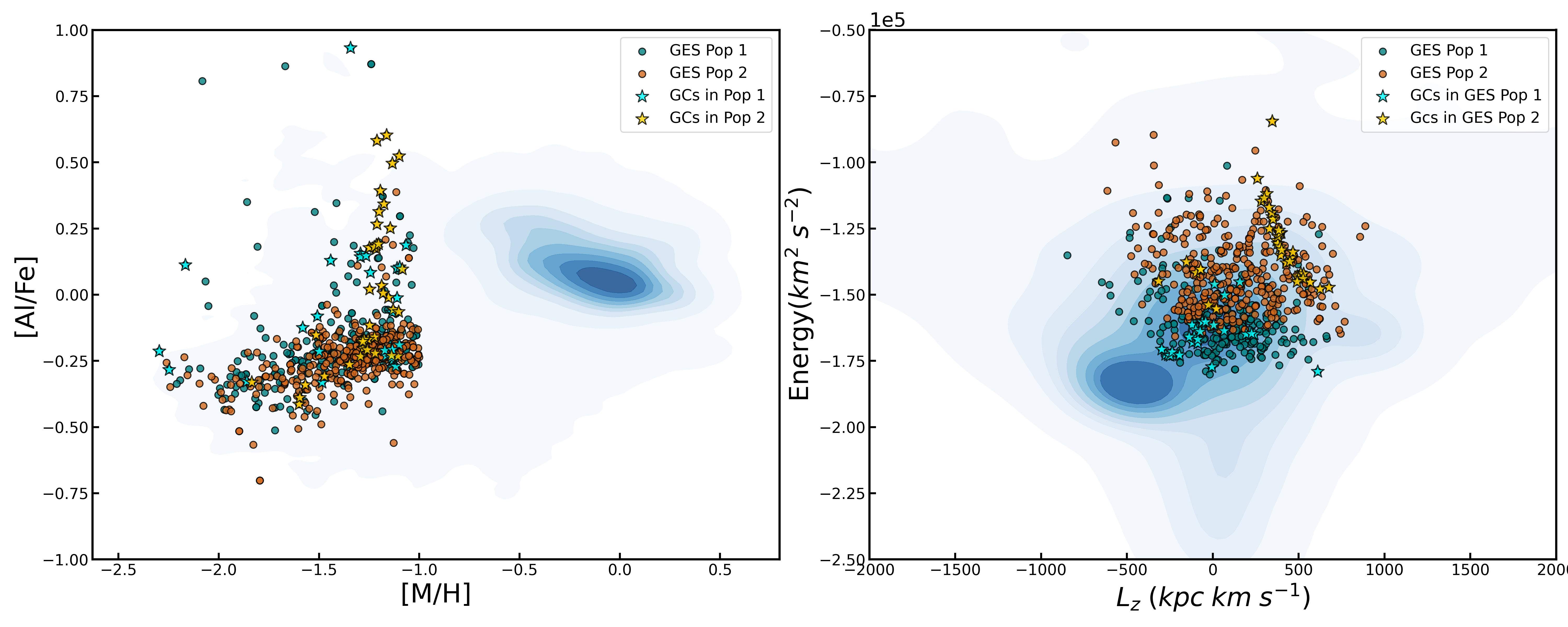}
    \caption{Left panel: [Al/Fe] versus [M/H] for the whole APOGEE dataset, inclusive of disc (blue contours). Pop~1 and Pop~2 from this work are highlighted in green and orange and individual stars in globular clusters are represented as light blue and yellow stars for the two populations respectively. Right panel: Energy versus angular momentum for the APOGEE halo (blue contours). Pop~1 and Pop~2 from this work are highlighted in green and orange and individual stars in globular clusters are represented as light blue and yellow stars for the two populations respectively.}
    \label{fig:Al}
\end{figure*}

GES consists of a significant fraction of the halo stars in our sample. Interestingly, our selection includes the literature one, as that of \citet{Horta2022}, but it is even broader, yielding a distinct sample of GES stars that appears to separate into two substructures. We refer to these as population one (Pop~1) and population two (Pop~2), associating Pop~1 with the low-energy, centrally concentrated, and in theory chemically richer population, and Pop~2 with the high-energy, more chemical poor population. The right panel of Fig.~\ref{fig:Al} illustrates the separation in orbital energy. 

\begin{figure*}
    \centering
    \includegraphics[width=1.0\linewidth]{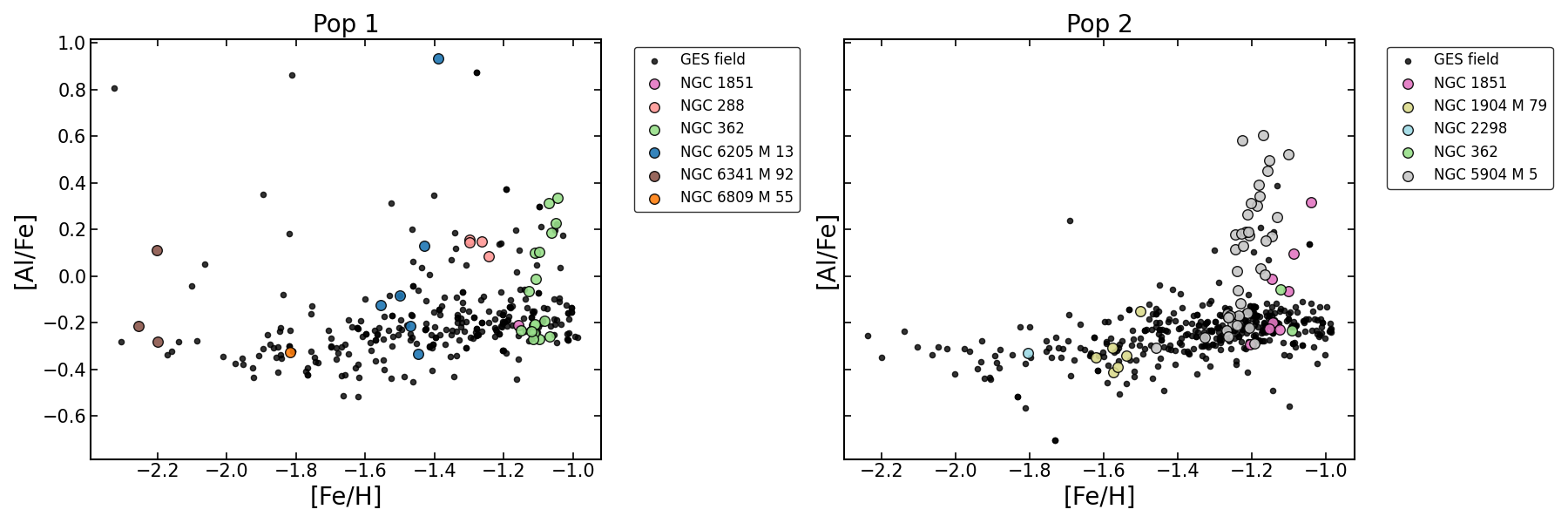}
    \caption{Left panel: [Al/Fe] versus [Fe/H] for Pop~1 of Gaia Enceladus. Globular clusters are colour-coded according to the legend. Right panel: [Al/Fe] versus [Fe/H] for Pop~2 of Gaia Enceladus. Globular clusters are colour-coded according to the legend}
    \label{fig:GESwithGCs}
\end{figure*}

These two populations include both field stars and globular clusters, which are both part of GES, for a total of 325 stars in Pop~1 and 445 stars in Pop~2. The properties of these populations -- including ID, coordinates, orbital properties and abundances -- are represented in Tables \ref{tab:Population1} and \ref{tab:Population2}.
As expected, the identified GES populations occupy the region associated with accreted systems, a conclusion supported by the low [Al/Fe] abundances of its member stars compared to MW stars (see left panel of Fig.~\ref{fig:Al}).
To obtain an unbiased characterisation of GES field population, we exclude stars classified as members of GCs. These stars indeed show distinct abundance patterns, such as Na-O anti-correlations and enhanced scatter in [Al/Fe] at certain metallicities (see Fig.~\ref{fig:GESwithGCs}). Although we do not analyse these GCs in detail here, it is noteworthy that, according to their dynamics, they are classified as originating from GES in \citet{Massari2019}. These results demonstrate that {\sc CREEK} effectively recovers the origin of the chemo-kinematic structures, linking them to their progenitor galaxy. 
Generally, GC members belong to either Pop~1 or Pop~2, with no cluster having stars in both. NGC 1851 is an exception, with most (eight) of its stars in Pop~2 and one star in Pop~1. 
Finally, there are few Al-rich stars that do not belong to any known GC, suggesting that their enrichment may instead result from mass-transfer binaries \citep{Usman2025}, other local enrichment processes \citep{BastianLardo2018}, or they were part of disrupted GCs \citep{Fernandez-Trincado2022}.

\setlength{\tabcolsep}{6pt}
\begin{table*}[t]
    \centering
    \small
    \begin{threeparttable}
    \caption{Summary of population 1 stars.}
    \begin{tabular}{llllllll}
    \hline
        APOGEE\_ID & ra & dec & $\cdots$ &[Ca/Fe] & [Fe/H] & [Ce/Fe]\\
        \hline
        2M22061609-3442491	& 22:06:16.09	& -34:42:49.24	& $\cdots$ &	0.20$\pm$0.03 &	-1.318$\pm$0.012 &	-0.09$\pm$0.04\\
        2M17081955+3405203 & 17:08:19.56 & +34:05:20.33 & $\cdots$ & 0.27$\pm$0.04 &	-1.524$\pm$0.013 &	-0.40$\pm$0.11\\
        2M16432684+2056177 & 16:43:26.83 & +20:56:17.68 & $\cdots$ & 0.11$\pm$0.03	& -1.565$\pm$0.012	& -0.24$\pm$0.05\\
        2M16475427-1902220 & 16:47:54.27 & -19:02:22.07 & $\cdots$ & 0.20$\pm$0.04	& -1.329$\pm$0.014	& -0.33$\pm$0.06\\
        2M16421482+3627199 & 16:42:14.81 & +36:27:19.85 & $\cdots$ & 0.10$\pm$0.04	& -1.499$\pm$0.013 &	-0.09$\pm$0.07\\
        2M03553890-0700174 & 03:55:38.91 & -07:00:17.71 & $\cdots$ & 0.20$\pm$0.03	& -1.286$\pm$0.011	& -0.21$\pm$0.06\\
    \hline
    \end{tabular}
    \begin{tablenotes}
    \small
    \item \textbf{Notes.} The complete Table will be available at the CDS.
    \end{tablenotes}
    \label{tab:Population1}
\end{threeparttable}
\end{table*}

\setlength{\tabcolsep}{6pt}
\begin{table*}[t]
    \centering
    \small
    \begin{threeparttable}
    \caption{Summary of population 2 stars.}
    \begin{tabular}{llllllll}
    \hline
        APOGEE\_ID & ra & dec & $\cdots$ &[Ca/Fe] & [Fe/H] & [Ce/Fe]\\
        \hline
        2M13335864+4410387	&13:33:58.62	&	+44:10:36.66	& $\cdots$ &	0.27$\pm$0.03 & -1.212$\pm$0.009 & -0.85$\pm$0.10\\
        2M15163358+0901284 & 15:16:33.58 & +09:01:28.31 & $\cdots$ & 0.24$\pm$0.03 & -1.248$\pm$0.011 &	-0.45$\pm$0.05\\
        2M10360156+3445298 & 10:36:01.57 & +34:45:29.84 & $\cdots$ & 0.61$\pm$0.06	& -2.138$\pm$0.016	& -0.38$\pm$0.08\\
        2M13393889+1836032 & 13:39:38.90 & +18:36:02.46 & $\cdots$ &0.22$\pm$0.03	& -1.285$\pm$0.010 &	1.68$\pm$0.07\\
        2M15212318-0109507 & 15:21:23.16 & +01:09:50.92 & $\cdots$ & 0.22$\pm$0.05	& -1.259$\pm$0.014	& -0.89$\pm$0.13\\
        2M16063833+2628574 & 16:06:38.33 & +26:28:57.38 & $\cdots$ & 0.18$\pm$0.03	& -1.166$\pm$0.014	& -0.05$\pm$0.06\\
    \hline
    \end{tabular}
    \begin{tablenotes}
    \small
    \item \textbf{Notes.} The complete Table will be available at the CDS.
    \end{tablenotes}
    \label{tab:Population2}
\end{threeparttable}
\end{table*}

\subsection{Chemical characterisation and statistical tests}

The two identified populations occupy distinct regions of the Lindblad diagram, as shown in the right panel of Fig.~\ref{fig:Al}.
Pop~1 corresponds to lower energies, while Pop~2 is found at higher energies. In the following, we provide evidence that Pop~1 might correspond to the inner, later-accreted component, whereas Pop~2 might represent the outer, earlier-accreted component of GES. The higher-energy population largely overlaps with previous GES selections from the literature, whereas the lower-energy population shows less overlap and may represent a previously unrecognized component.

The two substructures identified in GES are characterised by different energies, but also different chemical abundances.
In this section, we aim to give a chemical characterisation of these two populations.

\begin{figure*}
   \centering
   \small
   \includegraphics[width=1\linewidth]{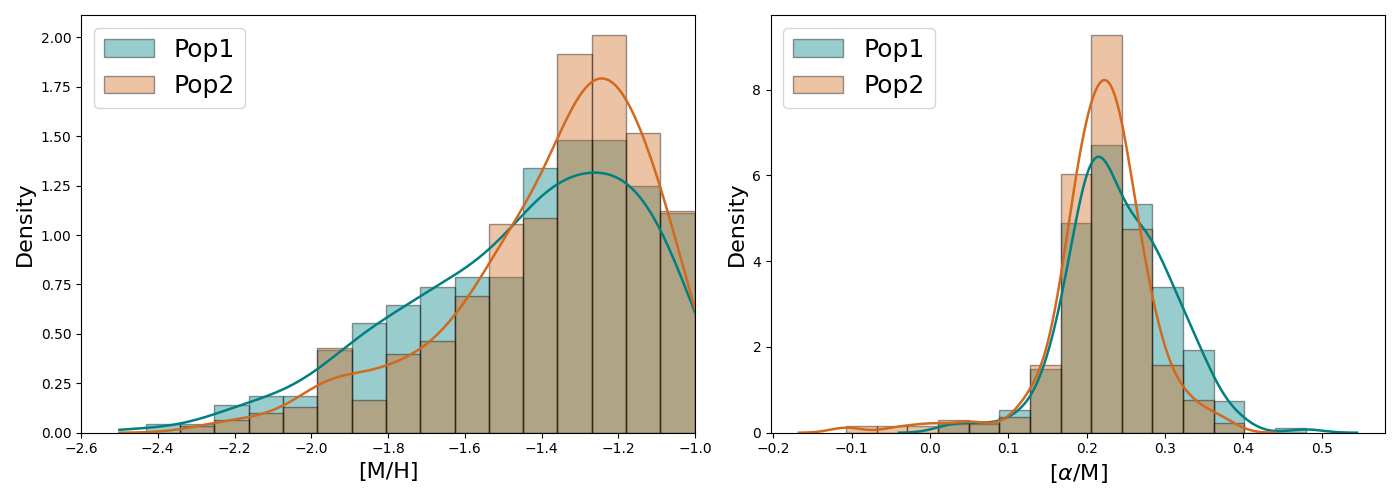}
   \caption{Left panel: Metallicity distribution function (MDF) for the two populations of Gaia-Enceladus. Right panel: histogram of [$\alpha$/M] for the two populations of Gaia-Enceladus. Continuous lines represent KDEs for the two distributions.} 
   \label{fig:mdf}
\end{figure*}

Different elements present in stellar atmospheres are produced though different nucleosynthetic processes. The so-called $\alpha$-elements (e.g. O, Mg, Si, Ca) are released into the interstellar medium predominantly by supernovae type II, at the end of the life of massive stars. In contrast, iron-peak elements (e.g. Cr, Fe, Ni) are mainly produced by supernovae type Ia, which occur in binary systems on longer timescales \citep[see e.g.][]{Nomoto13, Koba20SNIa}. Neutron-capture elements form via two primary channels: the slow (s-) process mostly in asymptotic giant branch (AGB) stars (e.g., Sr, Y, Ce), and the rapid (r-) process in explosive events such as neutron star mergers (e.g., Eu, Th). Consequently, variations in elemental abundances are expected among stellar populations that experienced different star formation histories \citep{Matteucci12}. The two populations analysed in this work both belong to GES, likely to two different regions, so the expected differences are relatively small but still statistically detectable.
To avoid potential biases driven by the particular data binning choices, in Fig.~\ref{fig:mdf}, the distribution of metallicity and $\alpha$-elements of the two populations are represented with their Kernel Density Estimation (KDE), a non-parametric method to estimate the underlying probability density function of the data. We see that, although the peaks occur at very similar positions, Pop~1 contains a larger fraction of metal-poor stars ([M/H] < –1.5) and shows a broader distribution around the peak, resulting in a lower maximum density. The KDE for $\alpha$-elements follows the same trend, with a wider distribution for Pop~1, in this case slightly shifted toward higher [$\alpha$/M].

\begin{figure*}
   \centering
   \includegraphics[width=0.9\linewidth]{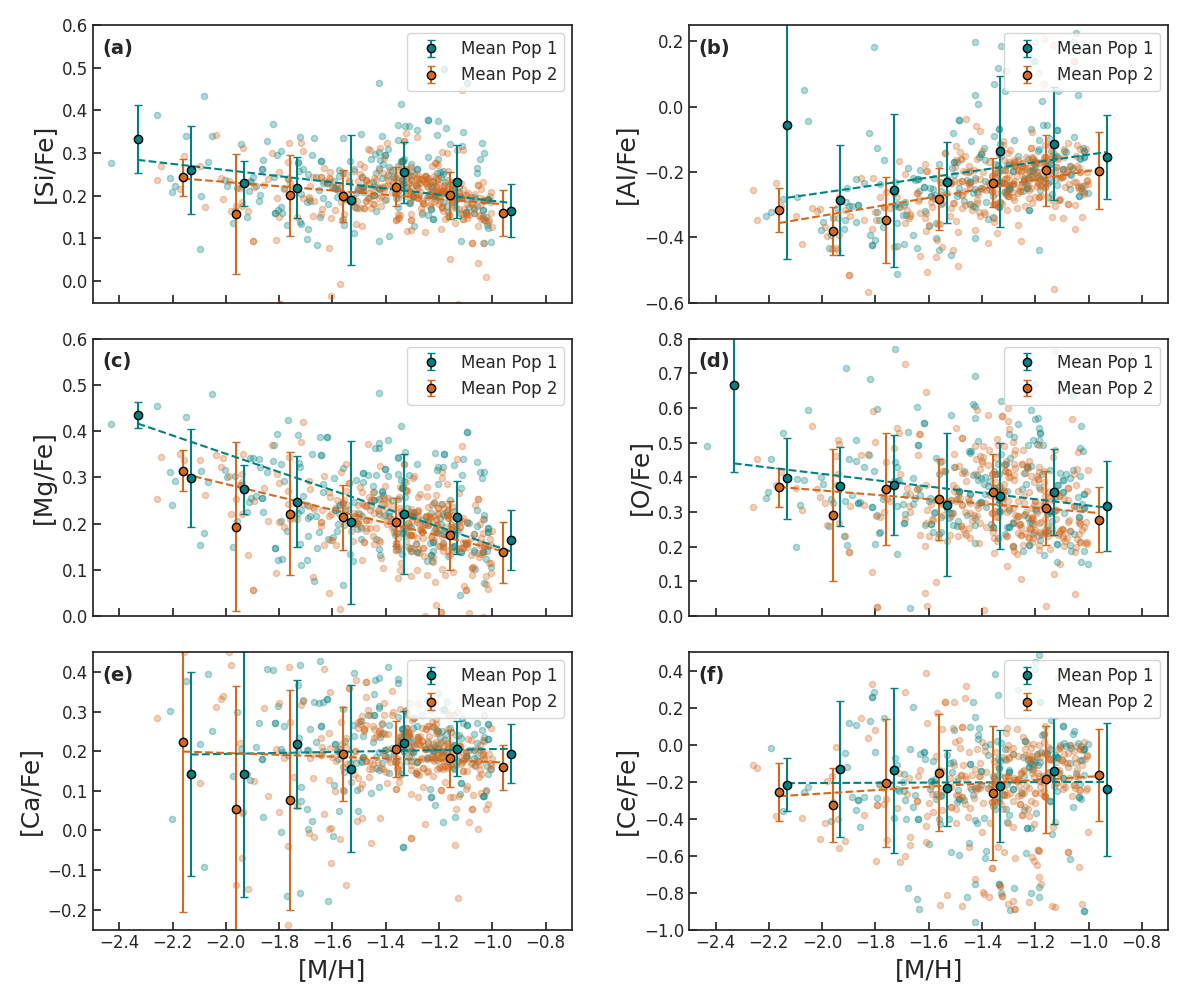}
   \caption{Panels a to f show [Si/Fe], [Al/Fe], [Mg/Fe], [O/Fe], [Ca/Fe], and [Ce/Fe] as a function of [M/H] respectively. The green and orange points correspond to Pop~1 and Pop~2, while the dashed lines indicate the linear fits to the mean abundance values in bins of metallicity.}
   \label{fig:density_all}
\end{figure*}

The distributions of other elements are superimposed at first glance, but their differences become significant in a statistical context.
By averaging abundances in bins of metallicity, we obtain the trends for several elements, as shown in Fig.~\ref{fig:density_all}. Exactly as expected for Pop~1, the abundances of Si, Al, Mg, O and Ca reflect a lower energy population characterised by slightly higher [X/Fe] abundance ratios and therefore higher star formation. Although the non-negligible abundance spread significantly hide the separation between the two populations, this remains coherent and consistent across all the mentioned panels.
We note that Ca does not show a clear separation between the two populations over the investigated  metallicity range. This is likely expected given the nucleosynthetic origin of Ca, due both to core-collapse supernovae and supernovae type Ia. Consequently, Ca behaves in an intermediate way between $\alpha$-elements and iron-peak elements.
For Ce the difference is even less pronounced. While the low-energy population still shows higher abundances compared to the high-energy population, the overlap between the two is substantial. However, Ce abundances should be treated cautiously, since APOGEE measurements are less reliable at low metallicity \citep[see e.g.][]{Cunha2017}.

To assess whether the two populations are chemically distinct in a statistical sense, we apply several non-parametric tests designed to evaluate differences in both central tendency and overall distribution. 
These non-parametric tests are robust and applicable even when the assumption of equal variances is violated.
The two sample Kolmogorov-Smirnov test \citep[K--S;][]{Kolmogorovan1933sulla}, the Cramér-von Mises \citep[CvM;][]{Anderson1952}, the Cucconi test \citep{Cucconi1968} and the energy distance test \citep{szekely2007} are all employed and provide similar results, confirming significant differences, namely very low p-values, in the elements Si, Al, Mg, O, and Ca. The energy distance test, implemented via permutation methods, is particularly robust and effective, especially in the presence of ties (identical values), where the K–S test may perform poorly.
For what concerns Ce, the K-S test returned a p-value below 0.05, even though in the other tests (CvM, Cucconi, energy distance) is quite above such a threshold, therefore indicating no significant difference between the populations. Given the known limitations of the K–S test with tied data, the energy distance test is considered more reliable.
The lack of statistically significant differences in neutron-capture elements as Ce, being predominantly produced by low-mass (< 3 M$_\odot$) stars, aligns with theoretical expectations, as its production takes place on longer timescales \citep{Chieffi2013}. The results are summarised in Table \ref{tab:stats_tests}.

\begin{table}[h]
    \centering
    \caption{P-values from statistical tests on different abundances: Kolmogorov–Smirnov (K--S), Cramér–von Mises (CvM), Cucconi, and Energy Distance respectively.}
    \begin{tabular}{lllll}
         \hline
         Abundance & K--S & CvM & Cucconi & Energy Distance \\
         \hline
         {[M/H]} & 0.004 & 0.004 & 0.000 & 0.000 \\
         {[Si/Fe]} & 0.000 & 0.000 & 0.000 & 0.000 \\
         {[Al/Fe]} & 0.001 & 0.001 & 0.000 & 0.000 \\
         {[Mg/Fe]} & 0.000 & 0.000 & 0.000 & 0.000 \\
         {[O/Fe]}  & 0.117 & 0.03 & 0.050 & 0.025 \\
         {[Ca/Fe]} & 0.010 & 0.019 & 0.031 & 0.061\\
         {[Ce/Fe]} & 0.0169 & 0.068 & 0.055 & 0.115 \\
         \hline
    \end{tabular}
    \label{tab:stats_tests}
\end{table}

\section{Origin of the two populations: insights from chemical evolution models}
\label{sec:models}

The comparison of distribution functions indicates that Pop~1 and Pop~2 experienced different enrichment pathways, likely reflecting differences in the star formation history. In the following, we compare the observed trends with predictions from galactic chemical evolution (GCE) models to better evaluate their origin.
These models describe the chemical enrichment in galaxies by means of analytical recipes (e.g. for gas flows, star formation) that allow to connect directly stellar abundance trends with the star formation history (SFH, see \citealt{Matteucci21} for a review).
In this work, we develop a simple framework where several models are run by varying key parameters to match the features of Pop~1 and Pop~2 identified in the previous Section. In the following, we provide a summary of the model ingredients and ranges of physical parameters adopted throughout this Section.

\subsection{Ingredients and parameters of GCE models}
\label{ss:GCE_description}

All the models adopted in this paper rest on the basic equations that describe the evolution of a given element $i$ (see, e.g. \citealt{Matteucci21}):
\begin{equation}
    \dot{G}_i (t) = -\psi(t)\, X_i(t)\, + \,R_i(t)\, +\, \dot{G}_{i,inf}(t)\, -\, \dot{G}_{i,out}(t),
    \label{eq:basic_chemevo}
\end{equation}
where $G_i(t)$ = $X_i(t)\ G(t)$ is the fraction of the gas mass in the form of an element $i$, $G(t)$ is the fractional gas mass and $X_i(t)$ the fraction in mass of a given element $i$.
This expression incorporates all the main processes affecting chemical enrichment: gas accretion ($\dot{G}_{i,inf}$), star formation ($\psi\, X_i$), large-scale outflows ($\dot{G}_{i,out}$), and stellar nucleosynthesis ($R_i$).

As commonly assumed in the literature \citep{Vincenzo19,Koba20SNIa}, primordial gas is accreted to the galaxy at an exponentially decreasing rate:
\begin{equation}
    \Dot{G}_{i,inf}(t) \propto X_{i,inf} \ e^{-t/\tau_{inf}},
    \label{eq:gas_infall}
\end{equation}
where $X_{i,inf}$ is the element $i$ mass fraction for primordial composition and  $\tau_{inf}$ represents the e-folding time for gas accretion.
The star formation rate (SFR) is implemented according to the Kennicutt-Schmidt law \citep{Kennicutt98}: 
\begin{equation}
    \psi(t)= \nu \, G(t),
    \label{eq:SFR}    
\end{equation}
with the star formation efficiency (SFE) $\nu$ as the control parameter that represents the SFR per unit mass of gas, which is the inverse of the gas depletion timescale (SFE $=1/\tau_{depl}$). 
As for the interaction with the Galaxy, the models for GES allow for large-scale gas loss through outflows,
assumed to be proportional to the SFR \citep{Matteucci12,Matteucci21}:
\begin{equation}
    \Dot{G}_{i,out}(t) \propto X_i \ \psi(t),
\end{equation}
with equal outflow mass loading across different chemical elements $i$.
The initial / starting time for such outflows set as an input parameter. This parameter identifies the time at which GES started significant interaction with the MW, leading to large amount of gas lost by the GES system.

To probe the characteristics of GES populations, here we vary the different physical parameter, namely the gas accretion timescale $\tau_{inf}$, the SFE $\nu$ and the initial time for large-scale outflows $t_{out}$. 
A summary of the ranges of parameter explored and the relative grid spacing is provided in Table \ref{tab:model_params}. Our choices on the probed ranges reflect the constraints we have on GES as well as on its interaction with the MW. As for example, we tested SFEs typical of dwarf galaxies, namely from 0.03 to 0.15 Gyr$^{-1}$. Also, for the initial times for the gas loss episode, we explored values between 1 and 4 Gyr, corresponding to ages between $\simeq$13 and $\lesssim$10 Gyr, namely within the time of merging completion.

It is worth noting from Table \ref{tab:model_params} that we adopt a relatively broad grid spacing. Indeed, due to the hard cut in stellar metallicity to [M/H]$< -1$ dex within the adopted APOGEE sample (see Section \ref{sec:dataset}), we do not aim to provide a detailed characterisation in model parameters for the two populations, as the lack of information on the high-metallicity tail of the metallicity distribution allows some degree of degeneracy between different models. 
Rather, here we want to provide useful insights on the origin of putative GES populations, by identifying whether the differences in the stellar distribution functions can be explained by tangible differences in the models setup.

\begin{table}[h]
    \centering
    \small
    \caption{Parameter ranges of values and grid spacings tested for the chemical evolution models adopted in this work.}
    \begin{tabular}{lll}
         \hline
         Model parameter & Range & Spacing  \\
         \hline
         {$\tau_{inf}$} (Gyr) & [0.5-5.5] & 1 \\
         {$\nu$} (Gyr$^{-1}$) & [0.03-0.15] & 0.03 \\
         {$t_{out}$} (Gyr) & [1-4] & 1  \\
         \hline
    \end{tabular}
    \label{tab:model_params}
\end{table}

For what concerns the stellar nucleosynthesis, the models relax the instantaneous recycling approximation, therefore allowing the different elements to be restored to the interstellar medium according to the lifetimes of their stellar progenitors. These are weighted according to the stellar initial mass function, for which we use the one by \citet{Kroupa93}, extensively used to model the evolution of the MW components and GES (e.g.,  \citealt{Romano05,Vincenzo19,Palla20}).
We adopt well-tested yield sets by \citet{Karakas10} for low- and intermediate-mass stars, \citet{Nomoto13} for massive stars/supernovae type II and \citet{Iwa99} for supernovae type Ia. For the latter, the delay-time-distribution by \citet[see also \citealt{Palla21} for details]{MatteucciRecchi01} is assumed. 

\begin{figure}
    \centering
    \includegraphics[width=1\columnwidth]{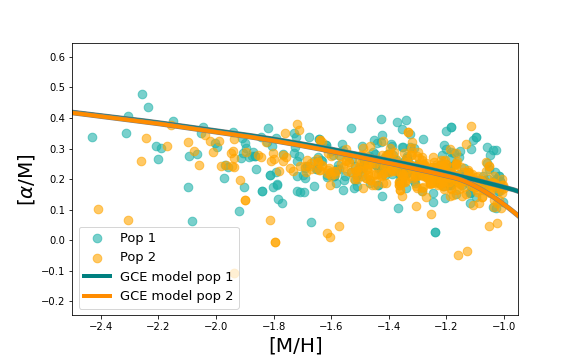}
    \caption{[$\alpha$/M] versus [M/H] for Pop~1 and Pop~2 as discussed in Section \ref{sec:dataset} (green and orange filled circles, respectively) as compared with predicted abundance ratios for the correspondent best models (green and orange solid lines, see Section \ref{ss:GCE_results}) .}
    \label{fig:alphaFe_model}
\end{figure}

Despite the importance of stellar yields in GCE models, these ingredients are chronically affected by relevant uncertainties (see thorough discussion in e.g. \citealt{Romano10,Prantzos18}), even on well-known $\alpha$-elements (see \citealt{Palla22}). This prevents us from looking into the details of the individual elements presented in Fig. \ref{fig:density_all}, as the underlying modelling uncertainties result to be much larger than the differences that can be detected between the samples.
Therefore, in the following discussion we focus on the metallicity and the global $\alpha$ abundance. It is worth noting that we lower in every run the predicted [$\alpha$/M]=[(Mg+Si+Ca+Ti)/M] by 0.05 dex, as the yield setup adopted results in a systematic offset of the observed [$\alpha$/M] versus [M/H] trends. This further motivates our approach to identify broad changes in the model physical parameters to explain the different GES populations rather than to provide a detailed tuning.
By applying the shift in [$\alpha$/M], we are able to nicely recover the abundance patterns as observed in GES populations, as shown in Fig. \ref{fig:alphaFe_model} for the two best models for Pop~1 and Pop~2 that are presented in the following Section.

\subsection{Comparison between observed and predicted populations}
\label{ss:GCE_results}

By correcting our model outputs for the adopted data selection function (see Section \ref{sec:dataset})\footnote{to obtain the selection function in terms of predicted ages and metallicities of the model, we adopt PARSEC evolutionary tracks (see \citealt{Palla22} for details of the procedure).} and comparing them with the data, we find that both Pop~1 and Pop~2 are well reproduced by models with rather small SFE ($\sim$0.06 Gyr$^{-1}$, namely more than a factor 10 less than typically assumed for the Galaxy) and initial time for outflows of 3 Gyr. Considering an age of the Universe of $\sim 13.8$ Gyr \citep{Planck18}, this translates in the main interaction starting between 11 and 10 Gyr ago, in line with previous results (e.g. \citealt{Helmi18,Montalban21,GonzalesKoda25}) identifying for the GES-MW merger an age of $\sim10$ Gyr.

The two best models that reproduce the observed distribution functions for the two populations as illustrated in Fig. \ref{fig:mdf} show very different gas accretion timescales. In particular, the model reproducing Pop~1 features adopts $\tau_{inf}=0.5$ Gyr, whereas the model for Pop~2 $\tau_{inf}=2.5$ Gyr. Such different gas accretion timescales are  consistent with an inside-out formation scenario for a proto-disc, where the inner components assemble on short timescales, while the outer components grow more slowly.
Despite not having the goal to provide detailed parameters tuning, the above mentioned SFE, $t_{out}$ and $\tau_{inf}$ setups give also best quantitive agreement (evaluated by means of the ${\rm L_2}$ metric) among those tested in this work with Pop~1 and Pop~2 MDFs and [$\alpha$/M]DFs.

\begin{figure*}
   \centering
   \includegraphics[width=1\linewidth]{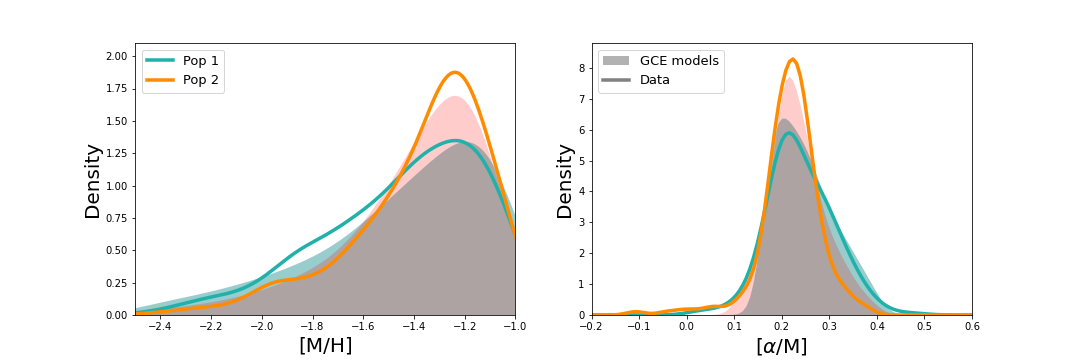}
   \caption{Kernel density estimations for observed stellar distribution functions for Pop~1 and Pop~2 as discussed in Section \ref{sec:dataset} (green and orange solid lines, respectively) as compared with predictions from the two correspondent best models (green and orange filled areas). Left panel: metallicity distribution function; right panel: [$\alpha$/M] distribution function.}
   \label{fig:mdf_model}
\end{figure*}

The KDE of the two best models compared with the observed MDFs and [$\alpha$/M]DFs are shown in Fig. \ref{fig:mdf_model}.
The best model for Pop~1 (hereafter, Pop~1 model) shows shallower distributions for both [M/H] and [$\alpha$/M] relative to the best model for Pop~2 (hereafter, Pop~2 model), in line with the observations.
Indeed, both the models well reproduce the respective distribution functions. Only Pop~1 model MDF show a KDE peak at [M/H]$\sim$ 0.04 dex larger than the observed one.
However, we note that the MDF of Pop~1 shows the largest sampling variability among the observed distributions, with a standard error in the KDE peak of $\simeq$0.04 dex (for the other distributions, standard errors are of the order or below 0.01 dex), which allows to reconcile observations with model predictions.
Moreover, we remind again that building models that reproduce the finer details of the distributions for the two identified GES populations is  beyond the scope of this paper.

Also, the imposed cut-off of stars at [M/H]$=-1$ dex likely affects the distributions and especially the shallower Pop~1 distribution, supposedly having a more significant contribution by "metal-rich" stars on the global stellar budget.
In any case, even testing a more stringent cut-off at [M/H]$=-1.1$ dex does not change the general picture: both the best models identified previously and the restricted samples display different behaviours for Pop~1 and Pop~2, with the former showing shallower distributions.

In general, the result obtained by the comparison between data and our simple model framework is remarkable.
The agreement for Pop~1 with models with much shorter gas accretion timescales than Pop~2 clearly points towards an inside-out scenario in the formation of the GES progenitor, similarly to what happens for the MW disc \citep{Matteucci89,Boissier99,Schonrich17}. Indeed, the high-energy Pop~2 likely denotes the outer regions of the GES progenitor, in contrast with the low-energy Pop~1, which is likely associated with the central regions (see also \citealt{Mori24,skuladottir2025evidence}). It is worth noting that such a result is obtained with no a priori assumptions in our models regarding the presence of a disc or an associated metallicity gradient in GES. 
Our results provide the first demonstration, based on chemical evolution models applied to GES data, of the inside-out nature of the relic galaxy. This strengthens the theoretical basis for the interpretation previously proposed by \citet{skuladottir2025evidence}, who identified a less chemically evolved, high-energy (outer) population and a more enriched, low-energy (inner) population in the Enceladus system.

\section{Summary and conclusions}
\label{sec:conclusions} 

The {\it Gaia} mission enabled the identification of multiple structures in the MW. GES represents one of the most significant accretion events in the history of the MW, corresponding to the merger of a massive dwarf galaxy that occurred about 10 Gyr ago. This event contributed a substantial fraction of the stellar halo, as well as several globular clusters, and is believed to have profoundly influenced the subsequent evolution of the Galaxy.
In this work, we analyse the two GES populations identified in \citet{berni2025exploring}. 
We obtained the following results:
\begin{itemize}
    \item Kinematically, Pop~1 occupies lower energies, while Pop~2 extends to higher energies typical of the more weakly bound, earlier-stripped material;
    \item Pop~1 shows a consistently slightly higher abundance and wider spread in all $\alpha$-elements available, verified with the application of statistical tests;
    \item GCE models suggest an inside-out formation scenario, without previous assumptions on metallicity gradients.
\end{itemize}

In conclusion, our analysis confirms that the two populations identified within GES are consistent with a multi-passage accretion event. During the first passage through the MW disc, the less bound, $\alpha$-depleted stars of the outer regions of the dwarf were stripped and deposited into the Galactic halo. In contrast, the second passage released the more tightly bound and $\alpha$-enhanced stars from the inner parts of GES. The chemical and dynamical signatures point toward an inside-out enrichment history for the progenitor galaxy. 
Future high-resolution multi-object spectrographs, such as HRMOS \citep{Magrini2023arXiv231208270M} and the Wide Spectroscopic telescope \citep[WST, ][]{Mainieri2024arXiv240305398M}, will allow an even more accurate chemical characterization of these stars, further advancing our understanding of accretion processes and the chemical evolution of progenitor systems.

\section*{Data availability}
Tables \ref{tab:Population1} and \ref{tab:Population2} are only available in electronic form at the CDS via \href{http://cdsweb.u-strasbg.fr/cgi-bin/qcat?J/A+A/}{http://cdsweb.u-strasbg.fr/cgi-bin/qcat?J/A+A/}.

\begin{acknowledgements}
We thank the referee for their careful reading of the manuscript and helpful suggestions. L.B. and L.M. acknowledge support from INAF through the Large Grants EPOCH and WST, funding for the WEAVE project, the Mini-Grants Checs (1.05.23.04.02), and financial support under the National Recovery and Resilience Plan (PNRR), Mission 4, Component 2, Investment 1.1, Call for tender No. 104 published on 2 February 2022 by the Italian Ministry of University and Research (MUR), funded by the European Union – NextGenerationEU, through the Project ‘Cosmic POT’ (Grant Assignment Decree No. 2022X4TM3H, MUR).
L.B., M.P., and L.M. acknowledge support from HORIZON-INFRA-2024-DEV-01-01 – Research Infrastructure Concept Development, through the project WST: The Wide-Field Spectroscopic Telescope (Grant No. 101183153).
M.P. also acknowledges support from the project "LEGO - Reconstructing the building blocks of the Galaxy by chemical tagging" granted by the Italian MUR through contract PRIN 2022LLP8TK\_001.
This work presents results from the European Space Agency (ESA)
space mission Gaia. Gaia data are processed by the Gaia Data Processing and Analysis Consortium (DPAC). Funding for the DPAC is provided by national institutions, in particular the institutions participating in the Gaia MultiLateral Agreement (MLA). The Gaia mission website is \url{https://www.cosmos.esa.int/gaia}. The Gaia archive website is \url{https://archives.esac.esa.int/gaia}.
\end{acknowledgements}

\bibliographystyle{aa}
\bibliography{aa58043-25}

\end{document}